\documentstyle[12pt]{article}
\begin{document}
\baselineskip=22pt plus 1pt minus 1pt

\begin{center}{\large \bf
The 3-dimensional $q$-deformed harmonic oscillator and magic numbers of 
alkali metal clusters}
\bigskip\bigskip

{Dennis Bonatsos$^{\#}$,
N. Karoussos$^{\#}$,
P. P. Raychev$^\dagger$,
R. P. Roussev$^\dagger$, 
P. A. Terziev$^\dagger$
\bigskip

{$^{\#}$ Institute of Nuclear Physics, N.C.S.R.
``Demokritos''}

{GR-15310 Aghia Paraskevi, Attiki, Greece}

{$^\dagger$ Institute for Nuclear Research and Nuclear Energy, Bulgarian
Academy of Sciences }

{72 Tzarigrad Road, BG-1784 Sofia, Bulgaria}}

\end{center}

\bigskip\bigskip
\centerline{\bf Abstract} \medskip
Magic numbers predicted by a 3-dimensional $q$-deformed harmonic oscillator 
with u$_q$(3) $\supset$ so$_q$(3) symmetry are compared to experimental 
data for alkali metal clusters, as well as to theoretical predictions 
of jellium models, Woods--Saxon and wine bottle potentials, and to the 
classification scheme using the $3n+l$ pseudo quantum number. The 
3-dimensional $q$-deformed harmonic oscillator correctly predicts all
experimentally observed magic numbers up to 1500 (which is the expected 
limit of validity for theories based on the filling of electronic shells), 
thus indicating that u$_q$(3), 
which is a nonlinear extension of the u(3) symmetry of the spherical 
(3-dimensional isotropic) harmonic oscillator, is a good candidate for 
being the symmetry of systems of alkali metal clusters.

\bigskip\bigskip
%PACS numbers: 33.10.Ev, 21.10.Re, 21.60.Ev

\newpage

Metal clusters have been recently the subject of many investigations
(see \cite{deHeer,Brack,Nester} for relevant reviews). One of the first 
fascinating findings 
in their study was the appearance of magic numbers 
\cite{Martin,Bjorn,Knight1,Knight2,Peder,Brec,Persson}, analogous to 
but different from the magic numbers appearing in the shell structure of 
atomic nuclei \cite{Mayer}. 
This analogy led to the early description of metal 
clusters in terms of the Nilsson--Clemenger model \cite{Clem},
which is a simplified version of the Nilsson model \cite{Nilsson1,Nilsson2} 
of atomic 
nuclei, in which no spin-orbit interaction is included. Further theoretical
investigations in terms of the jellium model \cite{Ekardt,Beck} 
demonstrated that the mean field potential in the case of simple metal 
clusters bears great similarities to the Woods--Saxon potential 
of atomic nuclei, with a slight modification of the ``wine bottle''
type \cite{Kotsos}. 
The Woods--Saxon potential itself looks like a harmonic 
oscillator truncated at a certain energy value and flattened at the bottom. 
It should also be recalled that an early schematic explanation of the 
magic numbers of metallic clusters has been given in terms of a scheme 
intermediate between the level scheme of the 3-dimensional harmonic 
oscillator and the square well \cite{deHeer}. Again in this case the 
intermediate 
potential resembles a harmonic oscillator flattened at the bottom.  

On the other hand, modified versions of harmonic oscillators 
\cite{Bie,Mac} have been 
recently investigated in the novel mathematical framework of quantum 
algebras \cite{Chari}, which are nonlinear generalizations of the usual Lie
algebras. The spectra of $q$-deformed oscillators increase either 
less rapidly (for $q$ being a phase factor, i.e. $q=e^{i\tau}$ with 
$\tau$ being real)
or more rapidly (for $q$ being real, i.e. $q=e^{\tau}$ with $\tau$ being 
real)
in comparison to the equidistant spectrum 
of the usual harmonic oscillator \cite{Roman}, while the corresponding 
(WKB-equivalent) potentials \cite{BDKJMP}
resemble the harmonic oscillator potential,
truncated at a certain energy (for $q$ being a phase factor) 
or not (for $q$ being real), 
the deformation inflicting an overall
widening or narrowing of the potential, depending on the value of the 
deformation parameter $q$.   

Very recently, a $q$-deformed version of the 3-dimensional harmonic 
oscillator has been constructed \cite{Terziev}, taking advantage of the 
u$_q$(3) $\supset$ so$_q$(3) symmetry 
\cite{Smirnov,Smirnov2,Smirnov3,Smirnov4,Jeugt,Jeugt2,Jeugt3}. 
The spectrum of this 3-dimensional $q$-deformed harmonic oscillator 
has been found \cite{Terziev} to reproduce very well the spectrum of the 
modified harmonic oscillator introduced by Nilsson 
\cite{Nilsson1,Nilsson2}, without the 
spin-orbit interaction term. Since the Nilsson model without the 
spin orbit term is essentially the Nilsson--Clemenger model used 
for the description of metallic clusters \cite{Clem}, it is worth examining 
if the 3-dimensional $q$-deformed harmonic oscillator can reproduce 
the magic numbers of simple metallic clusters. This is the subject 
of the present investigation. 

The space of the 3-dimensional $q$-deformed harmonic oscillator consists of 
the completely symmetric irreducible representations of the quantum algebra
u$_q$(3). In this space a deformed angular momentum algebra, so$_q$(3), 
can be defined \cite{Terziev}. 
The Hamiltonian of the 3-dimensional $q$-deformed 
harmonic oscillator is defined so that it satisfies the following 
requirements:

a) It is an so$_q$(3) scalar, i.e. the energy is simultaneously measurable
with the $q$-deformed  angular momentum related to the algebra so$_q$(3) 
and its $z$-projection.   

b) It conserves the number of bosons, in terms of which the quantum 
algebras u$_q$(3) and so$_q$(3) are realized. 

c) In the limit $q\to 1$ it is in agreement with the Hamiltonian of the usual 
3-dimensional harmonic oscillator. 
 
It has been proved \cite{Terziev} that the Hamiltonian of the 3-dimensional 
$q$-deformed harmonic oscillator satisfying the above requirements 
takes the form
\begin{equation}
H_q = \hbar \omega_0 \left\{ [N] q^{N+1} - {q(q-q^{-1})\over [2] } C_q^{(2)}
\right\},
\end{equation}
where $N$ is the number operator and $C_q^{(2)}$ is the second order 
Casimir operator of the algebra so$_q$(3), while 
\begin{equation}
[x]= {q^x-q^{-x} \over q-q^{-1}}
\end{equation} 
is the definition of $q$-numbers and $q$-operators. 

The energy eigenvalues of the 3-dimensional $q$-deformed harmonic oscillator 
are then \cite{Terziev}
\begin{equation}
E_q(n,l)= \hbar \omega_0 \left\{ [n] q^{n+1} - {q(q-q^{-1}) \over [2]}
[l] [l+1] \right\}, 
\end{equation}
where $n$ is the number of vibrational quanta and $l$ is the eigenvalue of the 
angular momentum, obtaining the values
$l=n, n-2, \ldots, 0$ or 1.  

In the limit of $q\to 1$ one obtains ${\rm lim}_{q\to 1} E_q(n,l)=
\hbar \omega_0 n$, which coincides with the classical result. 

For small values of the deformation parameter $\tau$ (where $q=e^{\tau}$)
one can expand eq. (3) in powers of $\tau$  obtaining \cite{Terziev}
$$
E_q(n,l)= \hbar \omega_0 n -\hbar \omega_0 \tau (l(l+1)-n(n+1))
$$
\begin{equation}
-\hbar \omega_0 \tau^2 \left( l(l+1)-{1\over 3} n(n+1)(2n+1) \right)
+ {\cal O} (\tau^3).
\end{equation}

The last expression to leading order bears great similarity to the modified 
harmonic 
oscillator suggested by Nilsson \cite{Nilsson1,Nilsson2} 
(with the spin-orbit term omitted)
\begin{equation}
V= {1 \over 2} \hbar \omega \rho^2 -\hbar \omega \kappa' 
({\bf L}^2 - <{\bf L}^2>_N ), \qquad \rho=r \sqrt {M\omega \over \hbar} ,
\end{equation}
where 
\begin{equation}
<{\bf L}^2>_N = {N(N+3)\over 2}.
\end{equation}
The energy eigenvalues of Nilsson's  modified harmonic oscillator
are \cite{Nilsson1,Nilsson2}
\begin{equation}
E_{nl}= \hbar \omega n -\hbar \omega \mu' \left( l(l+1)-{1\over 2}
n(n+3)\right). 
\end{equation}
It has been proved \cite{Terziev} that the spectrum of the 3-dimensional 
$q$-deformed harmonic oscillator closely reproduces the spectrum of 
the modified harmonic oscillator of Nilsson. In both cases the effect 
of the $l(l+1)$ term is to flatten the bottom of the harmonic 
oscillator potential, thus making it to resemble the Woods--Saxon 
potential. 

The level scheme of the 3-dimensional $q$-deformed harmonic oscillator 
(for $\hbar \omega_0 =1$ and $\tau = 0.038$) is given in Table 1, up to 
a certain energy. Each level is characterized by the quantum numbers 
$n$ (number of vibrational quanta) and $l$ (angular momentum). Next 
to each level its energy, the number of particles it can accommodate
(which is equal to $2(2l+1)$) and the total number of particles up to 
and including this level are given. If the energy difference between 
two successive levels is larger than 0.39, it is  considered as a gap 
separating two successive shells and the energy difference is reported 
between the two levels. In this way magic numbers can be easily read 
in the table: they are the numbers appearing above the gaps, written in 
boldface characters. 

The magic numbers provided by the 3-dimensional $q$-deformed harmonic 
oscillator in Table 1 are compared to available experimental data for 
Na clusters \cite{Martin,Bjorn,Knight1,Peder,Brec}
in Table 2 (columns 2--6). The following comments apply:

i) Only magic numbers up to 1500
are reported, since it is known that filling of electronic shells 
is expected to occur only up to this limit \cite{Martin}. For large 
clusters beyond this point it is known that magic numbers can be explained by
the completion of icosahedral or cuboctahedral shells of atoms \cite{Martin}. 

ii) Up to 600 particles there is consistency among the various experiments 
and between the experimental results in one hand and our findings in the 
other. 

iii) Beyond 600 particles the predictions of the three  experiments,
which report magic numbers in this region, are 
quite different. However, the predictions of all three  experiments are 
well accommodated by the present model. Magic numbers 694, 832, 1012
are supported by the findings of both Martin {\it et al.} \cite{Martin}
and Br\'echignac {\it et al.} \cite{Brec}, magic numbers 
1206, 1410 are in agreement with the experimental findings of Martin 
{\it et al.} \cite{Martin}, magic numbers 912, 1284 are supported by 
the findings of Br\'echignac {\it et al.}, 
while magic numbers 676, 1100, 1314, 1502 
are in agreement with the experimental findings of Pedersen {\it et al.}
\cite{Peder}. 

In Table 2 the predictions of three simple theoretical models \cite{Mayer}
(non-deformed 3-dimensional harmonic oscillator (column 9), 
square well potential  
(column 8), rounded square well potential (intermediate between the 
previous two, column 7)~) are also reported for comparison. It is clear 
that the predictions of the non-deformed 3-dimensional harmonic oscillator are 
in agreement with the experimental data only up to magic number 40, 
while the other two models give correctly a few more magic numbers (58, 
92, 138), although they already fail by predicting magic numbers at 68, 70, 
106, 112, 156, which are not observed.  

It should be noticed at this point that the first few magic numbers of 
alkali clusters (up to 92) can be correctly reproduced by the assumption 
of the formation of shells of atoms instead of shells of delocalized 
electrons \cite{Anagnos}, this assumption being applicable  under conditions 
not favoring delocalization of the valence electrons of alkali atoms. 

Comparisons among the present results, experimental data 
(by Martin {\it et al.} \cite{Martin} (column 2), Pedersen {\it et al.}
\cite {Peder} (column 3) and Br\'echignac {\it et al.} \cite{Brec}
(column 4)~) and 
theoretical predictions more sophisticated than these reported in Table 2,
can be made in Table 3, where magic numbers predicted by various 
jellium model calculations (columns 5--8, 
\cite{Martin,Bjorn,Brack,Bulgac}), Woods--Saxon 
and wine bottle potentials (column 9, \cite{Nishi}), as well as by a 
classification scheme using the $3n+l$ pseudo quantum number 
(column 10, \cite{Martin}) are reported. The following observations can be 
made:

i) All magic numbers predicted by the 3-dimensional $q$-deformed harmonic 
oscillator are supported by at least one experiment, with no exception.

ii) Some of the jellium models, as well as the $3n+l$ classification scheme, 
predict magic numbers at 186, 540/542, which are not supported by 
experiment. Some jellium models also predict a magic number at 
748 or 758, again without support from experiment. The Woods--Saxon 
and wine bottle potentials of Ref. \cite{Nishi} predict a magic number at 
68, for which no experimental support exists. The present scheme 
avoids problems at these numbers. It should be noticed, however, 
that in the cases of 186 and 542 the energy gap following them 
in the present scheme is 
0.329 and 0.325 respectively (see Table 1), i.e. quite close to 
the threshold of 0.39 which we have considered as the minimum energy 
gap separating different shells. One could therefore qualitatively 
remark that 186 and 542 are ``built in'' the present scheme as
``secondary'' (not very pronounced) magic numbers.  

The following general remarks can also be made:

i) It is quite remarkable that the 3-dimensional $q$-deformed harmonic 
oscillator reproduces the magic numbers at least as accurately as other,
more sophisticated, models by using only one free parameter ($q=e^{\tau}$). 
Once the parameter is fixed, the whole spectrum is fixed and no further 
manipulations can be made.
This can be considered as evidence that the 3-dimensional $q$-deformed 
harmonic oscillator owns a symmetry (the u$_q$(3) $\supset$ so$_q$(3)
symmetry) appropriate for the description of the physical systems under 
study. 

ii) It has been remarked \cite{Martin} that if $n$ is the number of nodes 
in the solution of the radial Schr\"odinger equation and $l$ is the 
angular momentum quantum number, then the degeneracy of energy levels of 
the hydrogen atom characterized by the same $n+l$ is due to the so(4) 
symmetry of this system, while the degeneracy of energy levels of the 
spherical harmonic oscillator (i.e. of the 3-dimensional isotropic 
harmonic oscillator) characterized by the same $2n+l$ 
is due to the su(3) symmetry of this system. $3n+l$ has been used 
\cite{Martin} to approximate the magic numbers of alkali metal clusters
with some success, but no relevant Lie symmetry could be determined
(see also \cite{Koch,Ostr}). In view
of the present findings the lack of Lie symmetry related to $3n+l$ is quite 
clear: the symmetry of the system appears to be a quantum algebraic 
symmetry (u$_q$(3)), which is a nonlinear extension of the Lie 
symmetry u(3). 

iii) An interesting problem is to determine a WKB-equivalent potential 
giving (within this approximation) the same spectrum as the 
3-dimensional $q$-deformed harmonic oscillator, using methods similar 
to these  of Ref. \cite{BDKJMP}. The similarity
between the results of the present model and these provided by the 
Woods--Saxon potential (column 9 in Table 3) suggests that the answer 
should be a harmonic oscillator potential flattened at the bottom, 
similar to the Woods--Saxon potential. If such a WKB-equivalent 
potential will show any similarity to a wine bottle shape,
as several potentials used for the description of metal clusters do
\cite{Ekardt,Beck,Kotsos},  remains to be seen. 

In summary, we have shown that the 3-dimensional $q$-deformed harmonic 
oscillator with u$_q$(3) $\supset$ so$_q$(3) symmetry correctly 
predicts all experimentally observed magic numbers of alkali metal clusters  
up to 1500, which is the expected limit of validity for theories based on 
the filling of electronic shells. This indicates that u$_q$(3), which 
is a nonlinear deformation of the u(3) symmetry of the spherical
(3-dimensional isotropic) harmonic oscillator, is a good candidate for 
being the symmetry of systems of alkali metal clusters.  

One of the authors (PPR) acknowledges support from the Bulgarian Ministry 
of Science and Education under contracts $\Phi$-415 and $\Phi$-547.

\parindent=0pt
 \newpage
%%%%%%%%%%%%%%%%%%%%%%%%%%%%%%%%%%%%%%%%%%%%%%%%%%%%%%%%%%%%%%%%%%%%%%
%%%%%%%%%%%%%%%%%%% Table 1 %%%%%%%%%%%%%%%%%%%%%%%%%%%%%%%%%%%%%%%%

Table 1: 
Energy spectrum, $E_q(n,l)$,  of the 3-dimensional $q$-deformed 
harmonic oscillator (eq. (3)), for $\hbar \omega_0 =1$ and 
$q=e^\tau$ with $\tau = 0.038$. Each level is characterized by $n$ 
(the number of vibrational quanta) and  $l$ (the angular momentum).
$2(2l+1)$ represents the number of particles each level can accommodate,
while under ``total'' the total number of particles up to and including 
this level is given. Magic numbers, reported in boldface, correspond to 
energy gaps larger than 0.39, reported between the relevant couples of 
energy levels.

\bigskip
\vfill\eject

\begin{table}

\caption{ }
\bigskip
\begin{tabular}{r r r r r || r r r r r}
\hline
$n$ & $l$ & $E_q(n,l)$ & $2(2l+1)$ & total & $n$ & $l$ & $E_q(n,l)$ & 
$2(2l+1)$ & total \\
\hline
 0&  0&  0.000 &  2  &  {\bf 2}  &    9&  5&  12.215 & 22&  462 \\
  &   &  1.000 &     &           &   11& 11&  12.315 & 46&  508 \\
 1&  1&  1.000 &  6  &  {\bf 8}  &   10&  8&  12.614 & 34&  542 \\
  &   &  1.006 &     &           &    9&  3&  12.939 & 14&  {\bf 556} \\
 2&  2&  2.006 & 10  & 18  &     &   &   0.397 &   &      \\
 2&  0&  2.243 &  2  & {\bf 20}  &    9&  1&  13.336 &  6&  562 \\
  &   &  0.780 &     &           &   12& 12&  13.721 & 50&  612 \\
 3&  3&  3.023 & 14  & {\bf 34}  &   10&  6&  13.863 & 26&  638 \\
  &   &  0.397 &     &           &   11&  9&  14.154 & 38&  {\bf 676} \\
 3&  1&  3.420 &  6  & {\bf 40}  &     &   &   0.603 &   &      \\
  &   &  0.638 &     &           &   10&  4&  14.757 & 18&  {\bf 694} \\
 4&  4&  4.058 & 18  & {\bf 58}  &     &   &   0.449 &   &      \\
  &   &  0.559 &     &           &   13& 13&  15.206 & 54&  748 \\
 4&  2&  4.617 & 10  & 68        &   10&  2&  15.316 & 10&  758 \\
 4&  0&  4.854 &  2  & 70        &   10&  0&  15.554 &  2&  760 \\
 5&  5&  5.116 & 22  & {\bf 92}  &   11&  7&  15.592 & 30&  790 \\
  &   &  0.724 &     &           &   12& 10&  15.777 & 42&  {\bf 832} \\
 5&  3&  5.841 & 14  &106        &     &   &   0.884 &   &      \\
 6&  6&  6.204 & 26  &132        &   11&  5&  16.660 & 22&  854 \\
 5&  1&  6.238 &  6  &{\bf 138}  &   14& 14&  16.779 & 58&  {\bf 912} \\
  &   &  0.860 &     &           &     &   &   0.606 &   &      \\
 6&  4&  7.098 & 18  &156        &   11&  3&  17.385 & 14&  926 \\
 7&  7&  7.328 & 30  &186        &   12&  8&  17.410 & 34&  960 \\
 6&  2&  7.657 & 10  &196        &   13& 11&  17.490 & 46& 1006 \\
 6&  0&  7.895 &  2  &{\bf 198}  &   11&  1&  17.782 &  6& {\bf 1012} \\
  &   &  0.502 &     &           &     &   &   0.667 &   &      \\
 7&  5&  8.396 & 22  &220        &   15& 15&  18.449 & 62& 1074 \\
 8&  8&  8.494 & 34  &{\bf 254}  &   12&  6&  18.660 & 26& {\bf 1100} \\
  &   &  0.627 &     &           &     &   &   0.645 &   &      \\
 7&  3&  9.121 & 14  &{\bf 268}  &   14& 12&  19.305 & 50& 1150 \\
  &   &  0.397 &     &           &   13&  9&  19.330 & 38& 1188 \\
 7&  1&  9.518 &  6  &274        &   12&  4&  19.554 & 18& {\bf 1206} \\
 9&  9&  9.709 & 38  &312        &     &   &   0.559 &   &      \\
 8&  6&  9.743 & 26  &{\bf 338}  &   12&  2&  20.113 & 10& 1216 \\
  &   &  0.894 &     &           &   16& 16&  20.226 & 66& 1282 \\
 8&  4& 10.637 & 18  &356        &   12&  0&  20.350 &  2& {\bf 1284} \\
10& 10& 10.980 & 42  &398        &     &   &   0.417 &   &      \\
 9&  7& 11.146 & 30  &428        &   13&  7&  20.767 & 30& {\bf 1314} \\
 8&  2& 11.196 & 10  &438        &     &   &   0.464 &   &      \\
 8&  0& 11.434 &  2  &{\bf 440}  &   15& 13&  21.231 & 54& 1368 \\
  &   &  0.781 &     &           &   14& 10&  21.360 & 42& {\bf 1410} \\
  &   &        &     &           &     &   &   0.475 &   &      \\
  &   &        &     &           &   13&  5&  21.835 & 22& 1432 \\
  &   &        &     &           &   17& 17&  22.119 & 70& {\bf 1502} \\
  &   &        &     &           &     &   &   0.441 &   &      \\
  &   &        &     &           &   13&  3&  22.560 & 14& 1516 \\
\hline
\end{tabular}
\end{table}

\newpage

%%%%%%%%%%%%%%%%%%% Table 2 %%%%%%%%%%%%%%%%%%%%%%%%%%%%%%%%%%%%%%%%

\begin{table}

\caption{ Magic numbers provided by the 3-dimensional $q$-deformed harmonic 
oscillator (Table 1), reported in column 1, are compared to the experimental 
data of Martin {\it et al.} [4] (column 2), 
Bj{\o}rnholm {\it et al.} [5] (column 3), 
Knight {\it et al.} [6] (column 4), 
Pedersen {\it et al.} [8] (column 5) and 
Br\'echignac {\it et al.} [9] (column 6), concerning Na clusters.
The magic numbers provided [11] by the (non-deformed) 3-dimensional harmonic 
oscillator (column 9), the square well potential (column 8) and a rounded
square well potential intermediate between the previous two (column 7) 
are also shown for comparison. See text for discussion. }

\bigskip

\begin{tabular}{c c c c c c c c c}
\hline
    &   exp.  & exp.    &   exp.   &  exp.& exp.  & int. & sq. well & h. osc.\\
%  our    Martin     Bjorn        Knight  Peder   Brec int    well        HO
 present & Ref. [4] &Ref. [5]& Ref. [6]& Ref. [8]& Ref. [9] & 
Ref. [11] & Ref. [11] & Ref. [11] \\  
\hline
    2  &    2      &   2     & 2&      &      &     2   &   2     &   2 \\
    8  &    8      &   8     & 8&      &      &     8   &   8     &   8 \\
  (18) &   18      &         &  &      &      &    18   &  18     &     \\
   20  &   20      &  20     &20&      &      &    20   &  20     &  20 \\
   34  &   34      &         &  &      &      &    34   &  34     &     \\
   40  &   40      &  40     &40&   40 &      &    40   &  40     &  40 \\
   58  &   58      &  58     &58&   58 &      &    58   &  58     &     \\
       &           &         &  &      &      &  68,70  &  68     &  70 \\
   92  &  90,92    &  92     &92&   92 &   93 &    92   & 90,92   &     \\
       &           &         &  &      &      &  106,112&  106    & 112 \\
  138  &  138      & 138     &  &  138 &  134 &    138  &  132,138&     \\
  198  &  198$\pm$2 & 196     & &  198 &  191 &    156  &   156   & 168 \\
  254  &           & 260$\pm$4& &      &      &         &         &     \\
  268  &  263$\pm$5 &         & &  264 &  262 &         &         &     \\
  338  & 341$\pm$5 & 344$\pm$4& &  344 &  342 &         &         &     \\
  440  & 443$\pm$5 & 440$\pm$2& &  442 &  442 &         &         &     \\
  556  & 557$\pm$5 & 558$\pm$8& &  554 &  552 &         &         &     \\
  676  &           &         &  &  680 &      &         &         &     \\
  694  &  700$\pm$15&         & &      &  695 &         &         &     \\
  832  &  840$\pm$15&         & &  800 &  822 &         &         &     \\
  912  &           &         &  &      &  902 &         &         &     \\
 1012  & 1040$\pm$20&         & &  970 & 1025 &         &         &     \\
 1100  &           &         &  & 1120 &      &         &         &     \\
 1206  & 1220$\pm$20&         & &      &      &         &         &     \\
 1284  &           &         &  &      & 1297 &         &         &     \\
 1314  &           &         &  & 1310 &      &         &         &     \\
 1410  &  1430$\pm$20&        & &      &      &         &         &     \\
 1502  &           &         &  & 1500 &      &         &         &     \\
\hline
\end{tabular}
\end{table}

\newpage
%%%%%%%%%%%%%%%%%%% Table 3 %%%%%%%%%%%%%%%%%%%%%%%%%%%%%%%%%%%%%%%%

\begin{table}

\caption{Magic numbers provided by the 3-dimensional $q$-deformed harmonic 
oscillator (Table 1), reported in column 1, are compared to the experimental 
data of Martin {\it et al.} [4] (column 2),
Pedersen {\it et al.} [8] (column 3), and 
Br\'echignac {\it et al.} [9] (column 4),
as well as to the theoretical predictions of various jellium model 
calculations reported by Martin {\it et al.} [4] (column 5), Bj{\o}rnholm 
{\it et al.} [5] (column 6), Brack [2] (column 7), Bulgac and Lewenkopf
[32] (column 8), 
the theoretical predictions of 
Woods--Saxon and wine bottle potentials reported by Nishioka {\it et al.}
[33] (column 9),
as well as to the magic numbers predicted by the  classification scheme 
using the $3n+l$ pseudo quantum number, reported by Martin {\it et al.}
[4] (column 10). See text for discussion.  
}

\bigskip

\begin{tabular}{c c c c c c c c c c}
\hline
         &exp. &exp. &exp. &jell.   & jell. &  jell.  & jell. & WS   &$3n+l$ \\
 present & Ref.[4] &Ref.[8]& Ref.[9]& Ref.[4]& Ref.[5] & Ref.[2] 
& Ref.[32]& Ref.[33]  & Ref.[4] \\  
\hline
%          exp   exp   exp jell        jell     jell          
%  our    Martin Ped  Brec Martin      Bjorn    Brack Bulgac  Nishi    3n+l 
  2 &    2     &     &     &  2     &    2 &     2   &       &     2  &    2 \\
  8 &    8     &     &     &  8     &    8 &     8   &       &     8  &    8 \\
(18)&   18     &     &     & 18     &   18 &         &       &        &   18 \\
 20 &   20     &     &     &(20)    &   20 &    20   &       &    20  &      \\
 34 &   34     &     &     & 34     &   34 &    34   &  34   &        &   34 \\
 40 &   40     &  40 &     &(40)    &   40 &         &       &    40  &      \\
 58 &   58     &  58 &     & 58     &   58 &    58   &  58   &    58  &   58 \\
    &          &     &     &        &      &         &       &    68  &      \\
 92 &  90,92   &  92 &  93 & 92     &   92 &    92   &  92   &    92  &   90 \\
138 &  138     & 138 & 134 &134     &  138 &   138   &  138  &   138  &  132 \\
    &          &     &     &186     &  186 &   186   &  186  &        &  186 \\
198 & 198$\pm$2& 198 & 191 &(196)   &  196 &         &       &   198  &      \\
254 &          &     &     &254     &  254 &   254   &  254  &   254  &  252 \\
268 & 263$\pm$5& 264 & 262 &(268)   &      &         &       &   268  &      \\
338 & 341$\pm$5& 344 & 342 &338(356)&  338 &   338   &  338  &   338  &  332 \\
440 & 443$\pm$5& 442 & 442 &440     &  440 & 438,440 &  440  &   440  &  428 \\
    &          &     &     &        &      &   542   &  542  &        &  540 \\
556 & 557$\pm$5& 554 & 552 &562     &  556 &   556   &  556  &   562  &      \\
676 &          & 680 &     &        &  676 &   676   &  676  &        &  670 \\
694 &700$\pm$15&     & 695 &704     &      &         &       &   694  &      \\
    &          &     &     &        &      &   758   &  748  &        &      \\
832 &840$\pm$15& 800 & 822 &852     &  832 &   832   &  832  &   832  &  820 \\
912 &          &     & 902 &        &      &   912   &  912  &        &      \\
1012&1040$\pm$20&970 &1025 &        &      &  1074   & 1074  &  1012  &  990 \\
1100&          &1120 &     &        &      &  1100   & 1100  &  1100  &      \\
1206&1220$\pm$20&    &     &        &      &         &       &  1216  & 1182 \\
1284&          &     &1297 &        &      &  1284   & 1284  &        &      \\
1314&          &1310 &     &        &      &         &       &  1314  &      \\
1410&1430$\pm$20&    &     &        &      &         &       &        & 1398 \\
1502&          &1500 &     &        &      &  1502   &  1502 &  1516  &      \\
 \hline
\end{tabular}
\end{table}

\end{document}